# Magnetically Superior and Easy to Handle $L1_0$-FePt Nanocrystals

Shinpei Yamamoto*, Yasumasa Morimoto, Teruo Ono and Mikio Takano

*Institute for Chemical Research, Kyoto University, Uji 611-0011, Japan*

Corresponding author

Address: Institute for Chemical Research, Kyoto University, Gokasyo, Uji, Kyoto 611-0011, Japan.

Tel.: +81-774-38-4718; fax: +81-774-38-3125

**E-mail address: <u>shinpei@msk.kuicr.kyoto-u.ac.jp</u> (S. Yamamoto)**

**Abstract**

We report a successful synthesis of monodisperse $L1_0$–FePt nanocrystals which are not only superior in magnetism but also easy to handle through being dispersible in solvents. Thanks to a thick $SiO_2$-coating, the thermal treatment needed to form the $L1_0$ structure can be done even at 900 $^oC$ without inducing coalescence and coarsening. The protecting shell was thereafter removed in a specific way that enabled us to recover the $L1_0$-FePt nanocrystals in a water-dispersion. The $SiO_2$-coated nanocrystals show a high coercivity of 18.5 kOe at room temperature in spite of their core size of only 6.5 nm in diameter, and the hysteresis loop of the water-dispersed nanocrystals, which were frozen under an external field, was almost rectangular indicating that their magnetic and structural orientation could be attained.

Arrays of monodisperse FePt nanocrystals in the $L1_0$ structure are excellent candidates for future recording media[1,2] with ultra-high densities beyond 1 Tb/inch$^2$. The most prominent feature of $L1_0$-FePt nanocrystals is that, for their strong magnetic anisotropy of *ca.* $10^8$ erg/cm$^3$, the superparamagnetic fluctuation of magnetization at room temperature can be suppressed down to a particle size of 2.8 nm in diameter[3]. So, the development of preparative methods that enable us to obtain highly crystalline, and therefore magnetically superior, monodisperse $L1_0$-FePt nanocrystals is a key demand. Another concern is that these particles should be easy to handle for the fabrication of desirable arrays on a substrate.

The synthesis of FePt nanoparticles through solution-based methods has been appreciated for the feasibility of morphplogical control[4-13]. However, only disordered fcc and $L1_0$ structures have been realized up to now, making a post thermal annealing necessary to transform them into the desired, well-ordered face-centered tetragonal (fct) form. Although the annealing temperature should be comparable to the thermodynamic Fe/Pt-ordering temperature of *ca.* 1300 °C, it is practically limited to 600 °C or below to prevent coalescence and coarsening of the nanoparticles. As an attempt to solve this problem, embedding of the fcc particles in a variety of chemically and thermally inert matrices has been examined[14-20]. However, it has been impossible to raise the annealing temperature high enough. Furthermore, these studies encountered difficulties in controlling the direction of their magnetic easy axis (parallel to the tetragonal *c* axis), which is extremely important from the viewpoint of practical use.

One of the key points of our synthetic strategy is the formation of a thick $SiO_2$-coating layer which acts as a "nanoreactor" as schematically illustrated in Fig. 1. Namely, the thermal diffusion of the Fe and Pt atoms during the heat treatment can be confined inside-the "$SiO_2$-nanoreactor". The other key point is the recovery of the thus formed $L1_0$-FePt nanocrystals in a solvent-dispersible manner for the ease of handling. Precursor fcc-FePt nanoparticles were prepared according to the method of Sun *et al*[4], and were subsequently coated by $SiO_2$ according to the method of Fan *et al*[21]. The particle composition was determined to be $Fe_{55}Pt_{45}$ by energy-dispersive X-ray analysis on a JEOL JED 2140. The $SiO_2$-coated fcc-FePt nanoparticles were annealed at various temperatures for 1 hr in flowing $H_2$(5 %)/Ar(95 %) gas to convert them to the $L1_0$ structure. Finally, the $SiO_2$ shell was dissolved in a tetramethylammonium (TMA) hydroxide solution (10 wt% aqueous solution) just by stirring the suspension at room temperature for 24 hrs.

Fig. 2 shows transmission electron microscopic (TEM) images of the nanoparticles in various synthetic stages taken by using a JEOL JEM-1010D[22]. The success in $SiO_2$-coating can be clearly seen in Fig. 2a, from which the average core size ($d$) and the standard deviation ($\sigma$) were estimated to be 6.4 nm and 15 %, respectively. Fig. 2b shows the particles annealed at 900 °C. The cores have remained the same in morphology and size. The $SiO_2$-free $L1_0$-FePt nanocrystals proved to be dispersible in water most probably because the surface was coated by TMA ions. A bottle of the aqueous solution is exhibited in the inset of Fig. 2c. Note that the dried nanocrystals were not coagulated as can be seen in Fig. 2c. There were no changes in shape and size upon the removal of the $SiO_2$ shell.

Structural features were examined by means of powder X-ray diffraction (XRD) using Cu $K_\alpha$ radiation ($\lambda$ = 0.154 nm) (Rigaku RINT2500). Fig. 3 shows a series of XRD patterns of the $SiO_2$-coated FePt nanoparticles in various preparative stages. The sample before the annealing shows three diffraction peaks characteristic of the disordered fcc-structure. The broad peak around $2\theta$ = 22 ° came from the amorphous $SiO_2$ shell. The conversion to the ordered $L1_0$ structure could be clearly seen only above 600~700 °C, though it was reported that bare nanoparticles began to transform below 600 °C[4]. The SiO$_2$-coating seems to have had a retarding effect as found for FePt-SiO$_2$ granular films[15]. The XRD patterns of the present samples annealed above 700 °C matched well that of $L1_0$-FePt. The degree of Fe/Pt ordering, which could be estimated from the relative intensity of the peaks representing the fcc and fct phases, was found to be improved up to 900 °C but no further above this temperature. After the dissolution of the SiO$_2$ shell, the broad peak around $2\theta = 22$ ° disappeared.

Magnetic properties were characterized by using a SQUID magnetometer (Quantum Design MPMS XL). Fig. 4a shows the room temperature hysteresis loop for the SiO$_2$-coated nanoparticles annealed at 900 °C. Here, $M_s$ represents the overall sample magnetization at 50 kOe. Because the amount of SiO$_2$ could not be determined precisely, it was impossible to make the magnetization specific with respect to the amount of the cores. It is notable that the coercivity, $H_c$, reaches a large value of 18.5 kOe in spite of the very small core size of 6.5 nm. This is the largest room temperature $H_c$ value ever reported for such small nanocrystals. The inset of Fig. 4a shows the variation of $H_c$ at 300 K as a function of the annealing temperature. Taking the XRD results into consideration, it is clear that $H_c$ sharply increases as the Fe/Pt-ordering is promoted. The fact that the ratio of $M_r / M_s = 0.5$ is also worth noting, where $M_r$ is the remnant magnetization. This indicates that the SiO$_2$-coated nanoparticles behave as an ensemble of randomly oriented, non-interacting, uniaxially anisotropic, single-domain particles because sintering of the SiO$_2$-shells and a tight core-shell adhesion prevent the orientation of the magnetic and structural axis along the external field even at 50 kOe.

The shape of the hysteresis loop changed drastically for the water-dispersed nanoparticles revealing that they are free from the above constraints. Fig. 4b shows the hysteresis loop of the water-dispersed nanocrystals measured after cooling the solution down to 200 K in an applied magnetic field of 50 kOe[23]. At this temperature, the matrix solution was frozen, suppressing any motion of the FePt nanocrystals. The shape is almost rectangular, with the $M_r$ being almost equal to the magnetization at ±50 kOe. This indicates that the magnetic easy axes of the present nanoparticles can be aligned parallel to the external field and that the magnetization can be reversed by applying an opposite field. The slight deviation from the ideal rectangular shape is most probably due to the presence of small particles having smaller magnetizations and/or smaller coercivities.

In conclusion, we have succeeded in preparing monodisperse, highly crystalline $L1_0$-FePt nanocrystals dispersible in water via the "$SiO_2$-nanoreactor" strategy. Their room temperature coercivity is as large as 18.5 kOe in spite of their small size of 6.5 nm. Such an extremely large $H_c$ value arises from the highly crystalline nature of the present particles attained by the annealing at sufficiently high temperatures. We also demonstrated that the solvent-dispersed $L1_0$-FePt nanocrystals orient their magnetic and structural axis along an external magnetic field. We expect that these nanocrystals can be made dispersible in various organic solvents also with the aid of proper surfactants. Such highly crystalline $L1_0$-FePt nanocrystals dispersible in various solvents are a promising material for the realization of ultra-high density recording.

23. The loop starts from ($M/M_s$, $H$) = (-1, -50 kOe), reaches the point of ($M/M_s$, $H$) = (0.6, 50 kOe), and comes back to the starting point.  The fact that the $M/M_s$ at the turning point of $H$= 50 kOe is considerably smaller than that at the starting point of $H$= -50 kOe seems to indicate a considerable distribution of $H_c$ around 35 kOe.  That is, nanoparticles having large coercivities do not reverse their magnetization even at $H$= 50 kOe.

**Acknowledgements**  The authors express their thanks to the MEXT, Japan for Grants-in-Aid of No. 12CE2005, No. 14204070, for COE Research on Elements Science, and for 21COE on Kyoto Alliance for Chemistry.

Figure captions

Figure 1 : Schematic illustration of synthesis of the $L1_0$-FePt nanocrystals dispersed in solvents via the "SiO$_2$-nanoreactor" strategy.

Figure 2 : TEM images of the SiO$_2$-coated FePt nanoparticles (a) before annealing, (b) after annealing at 900 °C and (c) after the removal of the SiO$_2$ shell. Inset shows an image of the solution containing the $L1_0$-FePt nanocrystals.

Figure 3 : XRD patterns of the SiO$_2$-coated FePt nanoparticles (a) before annealing and annealed at various temperatures of (b) 600 °C, (c) 700 °C, and (d) 900 °C, and (e) after the removal of the SiO$_2$ shell.

Figure 4 : (a) Room temperature hysteresis loop of the sample annealed at 900 °C. Inset shows plots of $H_c$ at 300 K as a function of annealing temperature. The solid line is a guide to eyes. (b) Hysteresis loop of the solvent-dispersed $L1_0$-FePt nanocrystals (annealed at 900 °C) measured at 200 K after cooling in an applied magnetic field of 50 kOe.

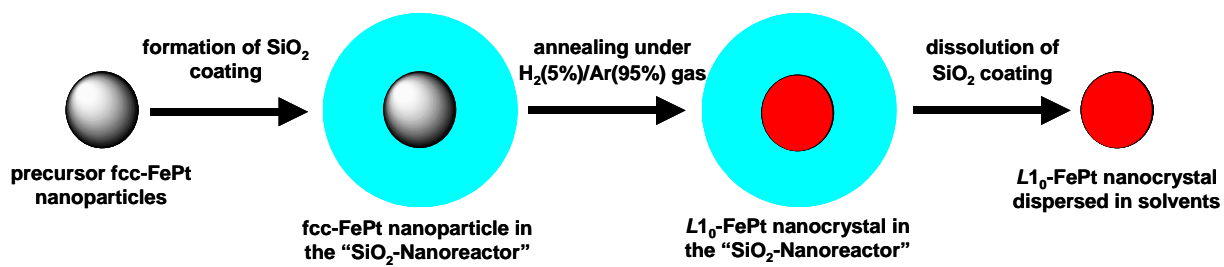

Figure 1

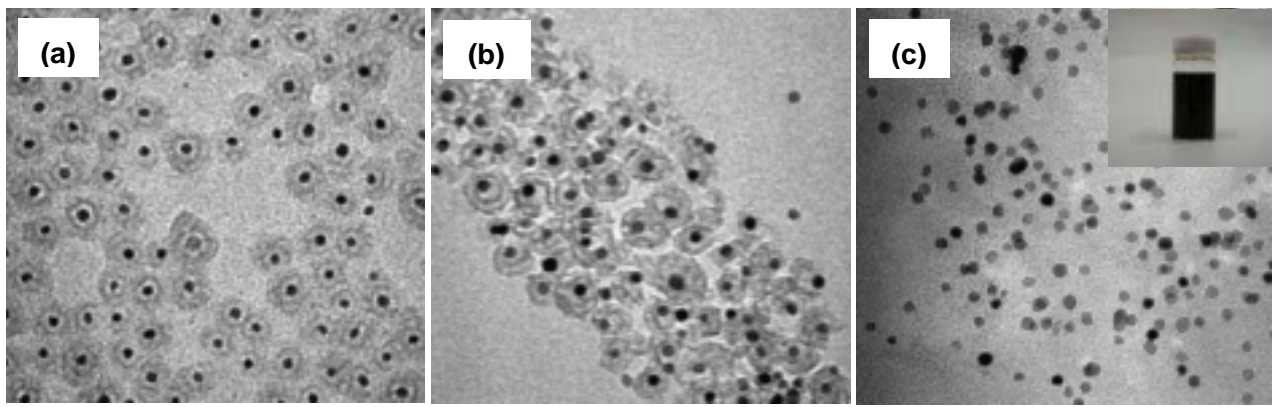

Figure 2

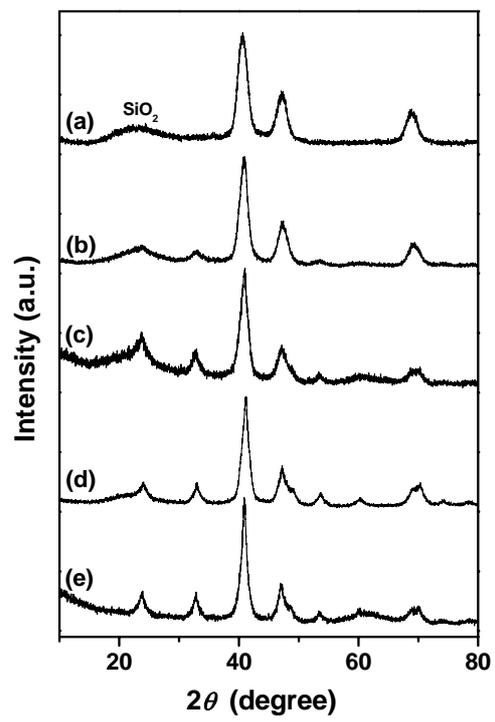

Figure 3

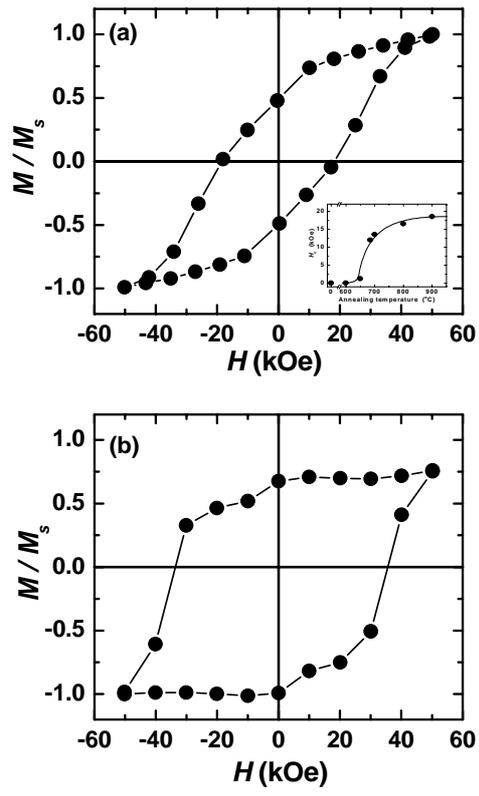

Figure 4